\documentclass[preprint]{aastex}

\newcommand{\lsim}{\mbox{\small$\stackrel{<}{\sim}$\normalsize}}   

\shorttitle{Flattening CCD Imaging Data}
\shortauthors{Marshall \& DePoy}
\begin{document}
\title{Flattening Scientific CCD Imaging Data with a Dome Flat Field System}
\author{J. L. Marshall \& D. L. DePoy}
\affil{Department of Astronomy, The Ohio State University}
\affil{140 West 18$^{th}$ Avenue, Columbus, Ohio 43210-1173\\
marshall@astronomy.ohio-state.edu, depoy@astronomy.ohio-state.edu}


\begin{abstract}
We describe the flattening of scientific CCD imaging data using a dome flat field system.  
The system uses light emitting diodes (LEDs) to illuminate a 
carefully constructed dome flat field screen. 
LEDs have several advantages over more traditional illumination sources: 
they are available in a wide range of output wavelengths, are inexpensive, have a very long 
source lifetime, and are straightforward to control digitally.  
The circular dome screen is made of a material with Lambertian scattering properties
that efficiently reflects light of a wide range of wavelengths and incident angles.  
We compare flat fields obtained using this new system with two types  
of traditionally-constructed flat fields: twilight sky flats and nighttime sky flats. 
Using photometric standard stars as illumination sources, we test
the quality of each flat field by applying it to a set of standard star observations.  
We find that the dome flat field system produces flat fields that are superior
to twilight or nighttime sky flats, particularly for photometric calibration.
We note that a ratio of the twilight sky flat to the nighttime sky flat 
is flat to within the expected uncertainty; but since both of these flat fields 
are inferior to the dome flat, this common test is not an appropriate metric for testing a flat field.
Rather, the only feasible and correct method for determining the appropriateness of a flat field is 
to use standard stars to measure the reproducibility of known magnitudes across the detector. 
\end{abstract}




\section{Introduction} \label{sec:intro}

Charge-coupled devices (CCDs) and other modern array detectors 
dominate optical and near-infrared observations in astronomy:
as of this writing there are nearly 6000 papers in refereed
journals with ``CCD'' in their titles or abstracts and many more
include infrared array data.
The primary reason for this ubiquity is 
that these detectors are nearly perfect for astronomical observations. 
They detect photons with high efficiency and produce an electronic signal 
capable of easy assimilation into digital recording, which has 
many subsequent advantages for data reduction and analysis. 
They also typically impose very little (often no significant) additional 
noise onto the observations. \cite{mclean} summarizes and discusses the 
history and advantages of array detectors in astronomy. 
New varieties of detectors are on the horizon \citep[e.g.][]{rando}, 
but CCDs and near-infrared arrays seem likely to continue to dominate 
ground-based optical and near-infrared observations
for many years.

CCDs and similar detectors have several characteristics that impose systematic 
signatures on any data they collect. 
The CCD's ``bias'' is a constant signal present in the absence of photons and 
independent of exposure time.   
``Dark current'' is not produced by photons, but does increase with 
exposure time.   
Some CCDs may not be very ``flat,'' i.e., they may have differences 
in the response to photons by the individual pixels 
of the detector \citep[although modern CCDs in particular can be very flat; see] 
[for an example]{morgan} and may require division by a ``flat field'' to remove these differences. 
Accurate interpretation of the data requires 
that these artifacts be removed during the reduction and analysis process.  
Bias and dark current are typically accounted for with special images taken without 
illumination of the detector, obtained under otherwise identical 
conditions (exposure time, device temperature, etc.) 
or with special electronic images obtained to mimic the signal present 
in a pixel without light (often added to the data frame and called 
an ``overscan'' region). There are other characteristics that can also 
affect data quality (non-linear response, charge transfer efficiency, 
residual signal, fringing, shutter timing, etc.), although these 
tend to be minor in modern detectors and for most 
applications \citep[see][for additional discussion]{mackay,mclean94}.

Unlike bias and dark current subtraction, flat field correction requires exposure to light and should
reproduce the illumination of the science observations onto the detector.
An ideal flat field removes the small response differences between 
pixels of the detector as well as field-dependent non-uniformities in the throughput of 
the optics. Unfortunately, a perfect flat field is difficult to obtain in practice for at least
two reasons. First, the detector must be illuminated in a manner 
that exactly replicates the illumination pattern of the actual observations. 
Necessarily this means that the pupil of the optics must be accurately reproduced 
or that the actual pupil must be evenly illuminated. The pupil can be defined 
as the largest optical surface that intercepts all angles from the observed field 
and is typically the primary mirror of the 
telescope \citep[although not always; see][]{zhou}. 
Achieving constant illumination over a large optic is challenging, 
particularly in the presence of scattered 
light \citep[see][]{gs96,kh99}. 
Second, the flatness of any detector depends on the spectrum of the 
incident light \citep{mackay}. Therefore, the illumination source must reproduce 
the color of the light that is to be detected. Reproducing this spectrum is 
generally impossible, since on any given image different sources have
different spectra and the sky will contribute a variable amount
depending on the relative brightness of the source.
The magnitude of the effect of a color mismatch between flat fields and observations 
when applying flat fields to CCD observations varies from detector to detector 
and is often not well measured.  

There are three common methods of obtaining a flat field. 
Most typically, images are obtained of either the bright twilight 
sky (``sky flats'') or of a nearby screen placed in front of the telescope that 
is illuminated by projected light (``dome flats''). Occasionally a flat 
field is made from the combination of many images taken of the nighttime sky. 
There are positive and negative aspects to each of these techniques. 
For example, twilight sky flats must be obtained at a relatively restricted 
range of times after sunset \citep[see][]{twilight} and can have significant 
gradients in illumination \citep[particularly for wide field observations; 
see][]{ch96}. 
In practice, it is often difficult to obtain all necessary
calibration images in the relatively short time available during twilight 
(particularly for inexperienced observers).
Dome flats are convenient to obtain and 
can have very high signal-to-noise ($S/N$), which reduces noise added to the science image by 
the data reduction process \citep[see][]{newb}. However, the color and pattern 
of the illumination may not reproduce the observations well. The presence of 
illumination sources not present during the actual observations (i.e. scattered 
light from the telescope or dome) can compromise the accuracy of all these 
techniques \citep[see][for example]{m01}.  Nighttime sky flats have other complications, 
mainly that it is difficult to obtain high enough $S/N$ for most applications.  
Other techniques for obtaining 
flat fields have been introduced: multiple exposures of time-independent 
signals at different spatial positions on the 
detector \citep{kll,tht03},
scanning extended sources \citep{dbw03}, dithered 
observations of non-uniform background sources \citep{wild}, and simultaneous 
observation of many photometrically-calibrated 
stars \citep{man95,man96}, for example. 
Most of these techniques have not been widely adopted at nighttime-oriented 
observatories.

In this paper we describe a dome flat field system that uses a 
specially constructed screen and banks of modern light 
emitting diodes (LEDs) as illumination sources. LEDs offer
significant advantages over more traditional sources of illumination and 
allow for the potential of greater flat field accuracy and improved 
photometric performance. We find that carefully constructed dome flats
outperform twilight or nighttime sky flats for photometric calibration of 
CCD images.


\section{Description of Dome Flat Field System}\label{sec:system}

Below we describe the two primary components of the dome flat field 
system installed at the McGraw-Hill 1.3 m telescope at the MDM
Observatory (http://www.astro.lsa.umich.edu/obs/mdm/).
The system consists of a screen and a set of illumination sources. The
screen is similar to that installed at many observatories, but it is 
nonetheless a crucial component of the system so we include
a description of its properties. The illumination system is a set
of LEDs that ring the top end of the telescope.

\subsection{Flat Field Screen}\label{sec:screen}

The flat field screen is a special piece of equipment;
it is not simply a white sheet draped against the dome wall.
The screen is circular in shape, stretched tightly onto
a metal ring to minimize wrinkles and shadows. 
The screen is attached to the interior of the dome
and mounted face-on to the secondary ring of the telescope.
Ideally, the screen should be perpendicular to the optical
axis of the telescope so that it best fills the aperture of the
telescope and provides the most predictable reflectance
pattern into the telescope beam. In practice, this geometry
is often difficult to obtain due to the mismatch between the 
locations of the center of rotation of the dome and telescope
axes. In Section~\ref{sec:characterization} we discuss the differences between dome flats
taken at a variety of telescope pointing and dome rotation 
combinations.

\begin{figure}
\plotone{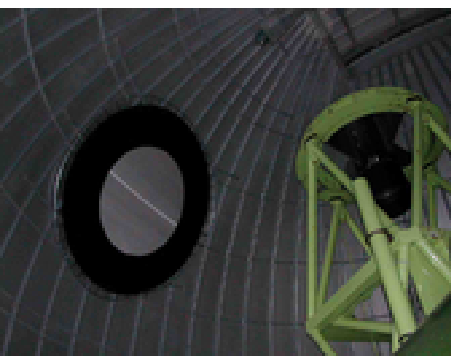}
\figcaption{
Photograph of the dome flat field screen at the McGraw-Hill 1.3m telescope.
The screen is attached to a tubular ring, which is mounted to the 
inside of the dome. The bright line is the seam between two panels of 
SORICSCREEN material, and has no measurable effect on the performance of the screen. 
Note the relatively large area of dark material around 
the white spot, which serves to reduce the amount of off-axis light 
scattered into the telescope beam.
The telecope is not pointed at the screen in this photograph.  
\label{fig:fig1}
}
\end{figure}

The screen consists of white fabric 
surrounded by a wide annulus of black fabric. 
Figure~\ref{fig:fig1} shows a picture of the screen mounted on the inside of the 
dome at the McGraw-Hill 1.3 m telescope. The diameter of the 
white area is 1.3 m; the surrounding black fabric is an annulus roughly 
0.6 m in width. The white spot is the same diameter as the primary mirror 
of the telescope so that it fills the pupil of the telescope 
well (but does not overfill it and thereby introduce extraneous off-axis 
illumination). The black material surrounding the reflective spot is essential
in reducing scattered light onto the primary 
mirror \citep[see][for a discussion of
scattered light issues]{gs96}.

\begin{figure}
\plotone{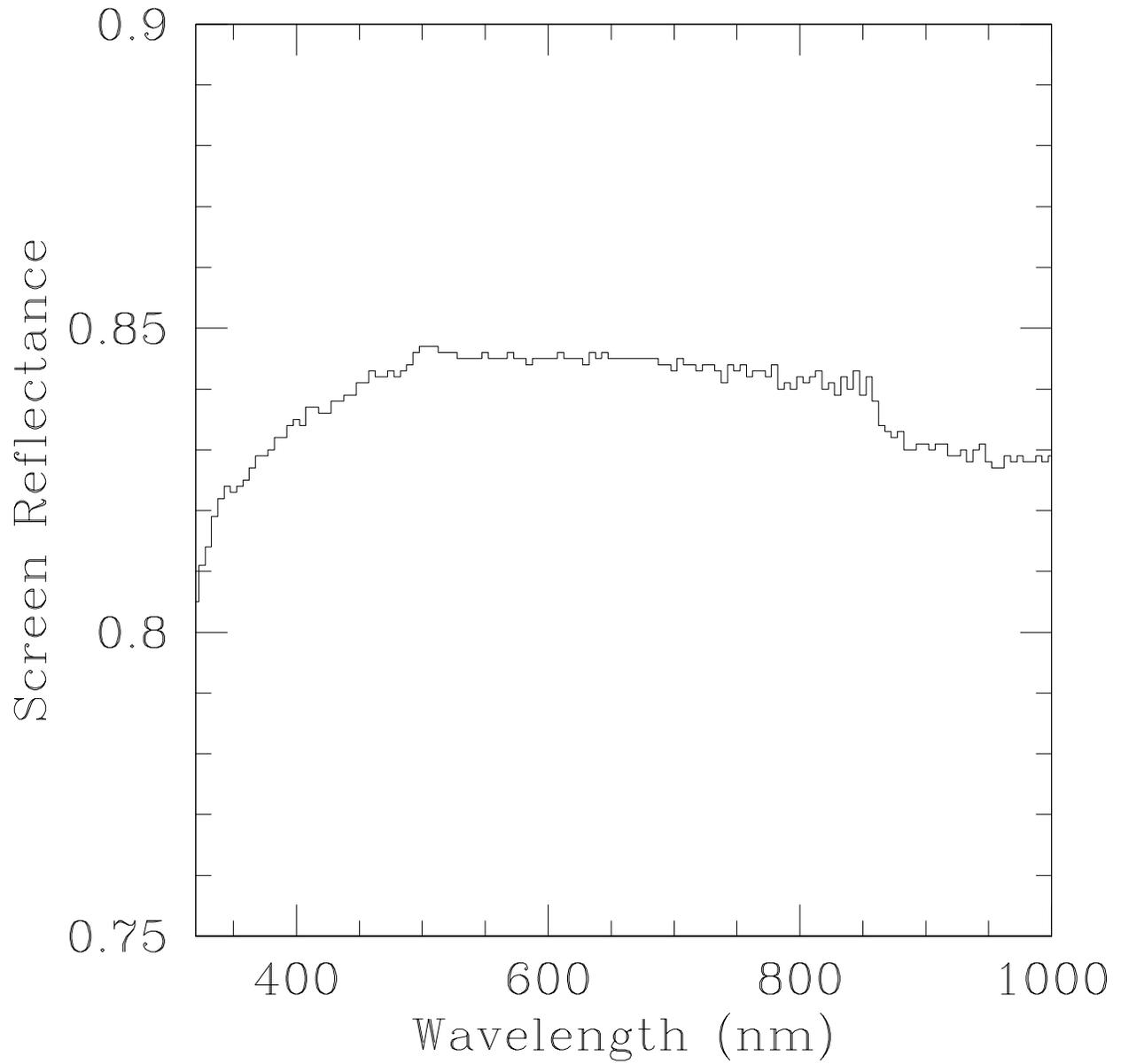}
\figcaption{
Reflectance of SORICSCREEN material in the ultraviolet to the 
near-infrared spectral region. The reflectance measurements were supplied by 
the screen manufacturer and are traceable to NIST standards.
The reflectance of the screen material is very constant throughout most of the 
optical wavelength range.  
\label{fig:fig2}
}
\end{figure}

The flat field screen was manufactured by Stellar Optics Research International 
Corporation of Thornhill, Ontario, Canada (http://www.soric.com). It is made of a 
durable ``SORICSCREEN'' fabric with high and relatively constant reflectance 
from the ultraviolet to the near-infrared. 
The screen has Lambertian scattering properties at a 
wide range of incident angles.
Figure~\ref{fig:fig2} shows the reflectance of the 
material versus wavelength, 
which varies only by $\sim$2\% from 350 nm to 1000 nm.
The material used for the white spot is more fully described by \cite{ms94}.

We note that this screen design is similar to that in use at the Cerro
Tololo Inter-American Observatory (CTIO) at
the Blanco 4 m telescope and other smaller telescopes. This system has
worked well for many years.

\subsection{Illumination Sources}\label{sec:leds}

\begin{figure}
\plotone{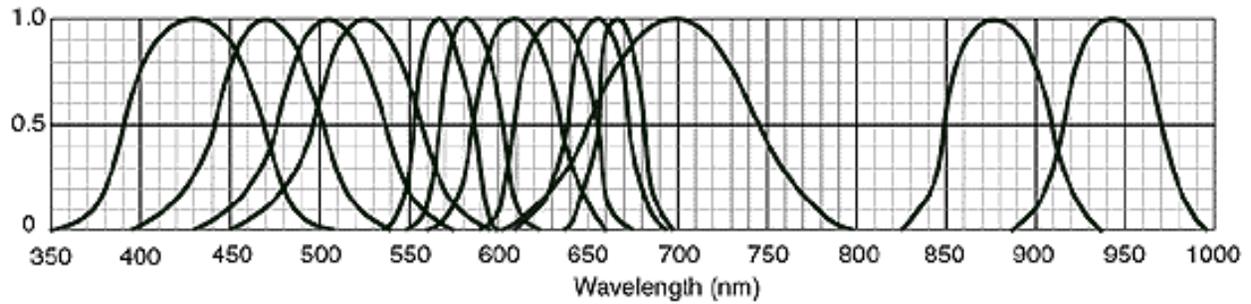}
\figcaption{
Spectral energy distributions of a sample of LEDs from Ledtronics, Inc.
The figure is taken from the Ledtronics web page (http://www.ledtronics.com).
Note that this is only a selection of the very wide range of colors available from this
single manufacturer; many more are available from other
vendors. Suitable combinations of these or other LEDs
can provide almost any illumination spectrum.
\label{fig:fig3}
}
\end{figure}

Light emitting diodes (LEDs) are very popular semiconductor devices that are found in an 
amazing variety of applications (e.g. flashlights, automobile taillights, 
supermarket scanners, and stadium picture displays). They are inexpensive 
(\$0.01 to \$10.00) and have extremely 
long lifetimes ($\sim$100,000 hours, depending on operating conditions). 
Furthermore, they can be controlled by simple digital electronics and 
require only low voltage for operation (typically 3-10 V). LEDs with very 
high brightness and an extremely wide range of colors have recently 
been developed. 
As an example, Figure~\ref{fig:fig3} shows the relative 
spectral energy distribution of the output light of
a number of LEDs from one particular manufacturer (Ledtronics, Inc.; see
http://www.ledtronics.com). There are many more LED 
manufacturers; for example, Roithner-Lazertechnik (http://www.roithner-laser.com) has a good
selection of far red LEDs that complement the bluer LEDs shown in Figure~\ref{fig:fig3}.
The combination of these characteristics suggest that LEDs 
make an excellent choice for flat field illumination sources.

\begin{figure}
\plotone{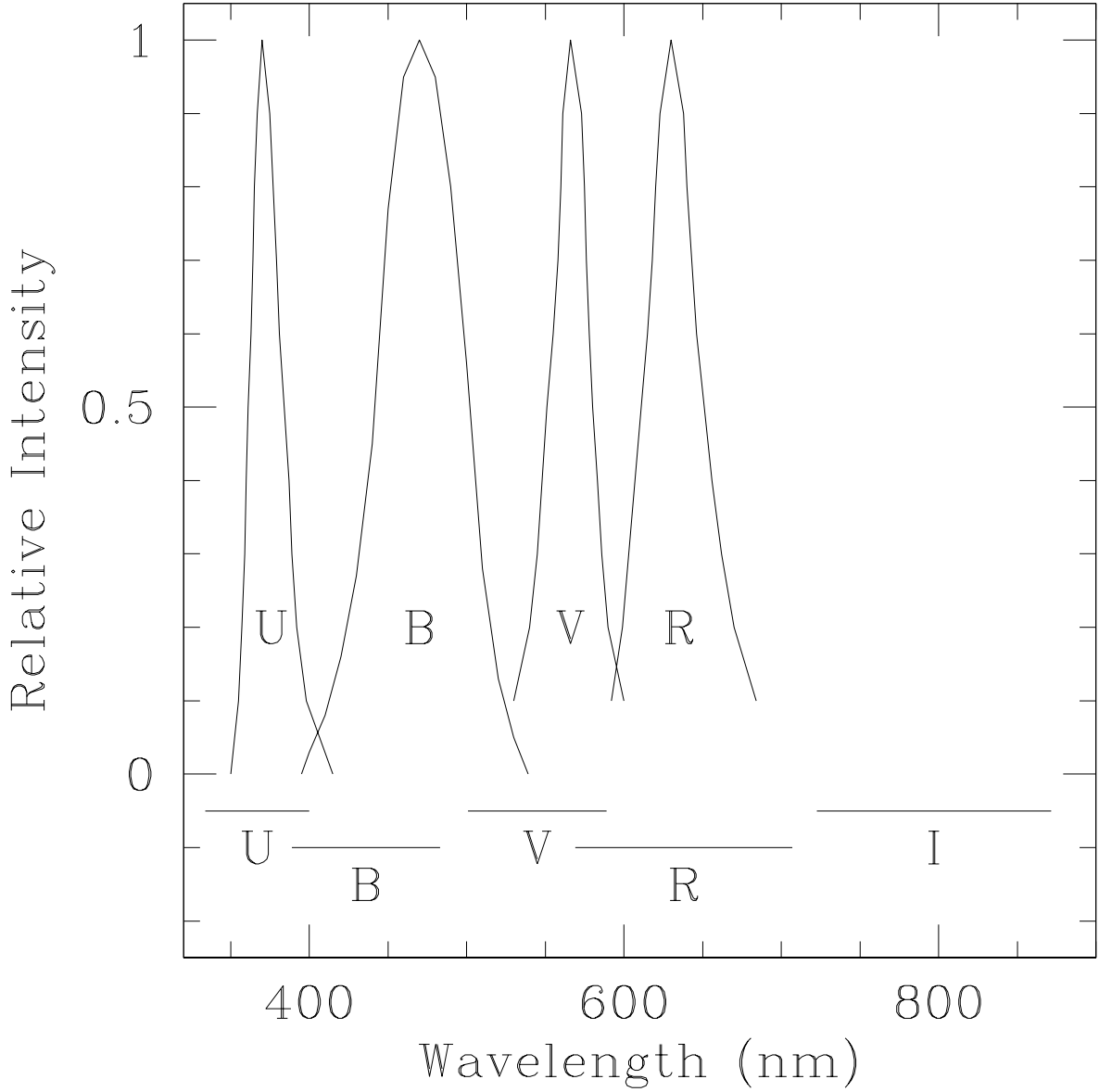}
\figcaption{
Spectral energy distributions of the UBVR LEDs used in the MDM 1.3m 
flat field system.  
Typical optical filter bandpasses are shown across the bottom of the figure.
These data are obtained from the manufacturers' 
websites. The equivalent data for the I band LED were not available to us 
(the manufacturer specifies $\lambda_C\approx810$ nm and 
$\Delta\lambda\approx35$ nm, however).
\label{fig:fig4}
}
\end{figure}

We selected five LEDs corresponding with each of the optical bandpasses (i.e., U, B, V, R, and I) 
in use at the telescope.  Each LED was chosen so that the central wavelength was close to
the center of the traditional UBVRI filters and with 
significant emission over most of the bandpass.  We refer to 
the LEDs by the bandpass with which they are associated.
We note that, to the eye, the color of the illumination looked approximately white 
with all the LEDs turned on.   
In general, the LEDs have relatively wide 
projection angles in order to illuminate the flat field screen fully. Table 
1 gives details of the specific LEDs we used. 
Figure~\ref{fig:fig4} shows the spectral energy distribution of the
light output of some of the LEDs used along with the
filter bandpasses (note that the brightness and peak output 
wavelength varies with temperature: 
$\sim1\%/^\circ$C in brightness and $\sim0.1$ nm $/ ^\circ$C in wavelength are typical).  
Again we emphasize that there are a 
wide variety of LEDs available with various colors; 
many combinations are possible that will suit other applications.
The LEDs are attached to the top of the secondary ring of the telescope. 
One LED of each of the five colors is mounted at each of the cardinal points of the ring. 

The most important consideration for the flat field screen illumination system 
is the uniformity of the radiance (radiant power from a given direction per unit 
area of the source per unit solid angle) 
of the screen as viewed from the focal plane of the telescope: 
the illumination of the focal plane must be uniform over the area of the 
field of view as well as the range of solid angles of incident light.  
We note that our screen is not very evenly illuminated (i.e., the screen looks mottled). 
While the illumination pattern on the screen does not appear very smooth to the eye, 
in fact, the combination of the scattering effects of the screen and the 
effects of the telescope optics produces a relatively uniformly 
illuminated field of view at the focal plane.  We demonstrate the 
uniformity of the radiance of our dome flat field system explicitly in 
Section~\ref{sec:characterization}. 

\section{Characterization of Performance of Flat Fields}\label{sec:characterization}

\subsection{Construction of the Flat Fields}\label{sec:construct}

We obtained twilight sky flats, dome flats, and nighttime sky flats at the 
MDM 1.3m telescope.  
Twilight sky flats were obtained in the usual manner \citep[see][]{twilight} 
during evening twilight under clear conditions.
We obtained dome flats at three different combinations
of telescope pointing and dome position.  These were 60 second exposures. 
We also created a flat field from
$\sim$75 long exposures of the nighttime sky. 
A large number of exposures were obtained to create each flat: approximately 20
images at each dome flat position, and 10 images of the twilight sky during one twilight period.
The typical signal was 15000-25000 electrons per pixel above the detector
bias level in each twilight sky and dome flat image, 
ensuring excellent formal $S/N$ in each combined flat ($S/N \sim$ 400. The
bias and any small dark signal was removed from the images prior
to combination into each final flat using standard IRAF
\footnote[1]{IRAF is distributed by the National Optical Astronomy Observatories, 
which is operated by the Association of Universities for Research in Astronomy, Inc., 
under cooperative agreement with the National Science Foundation.} utilities.

A 1024$\times$1024 SITe CCD
was used for all the measurements; the CCD has 0.024 mm pixels, which
correspond to $\sim$0.5 arcsec at the {\it f}/7.5 Cassegrain
focus of the 1.3 m telescope. 
A filter wheel mounted $\sim$50 mm above the CCD holds a Johnson V 
filter ($\lambda \approx$ 550 nm; $\Delta\lambda \approx$ 90 nm). All
measurements were made on 21 September 2003.
Care was taken to minimize effects not due to the illumination of the CCD
or its intrinsic properties. For example, exposure times for all flats were
sufficiently long that shutter timing effects are
small ($<$0.002 mag). Also, the maximum signal on the CCD was 
kept well below saturation and any non-linearity or charge
transfer effects should be $<$0.001 mag. 

\subsection{Direct Comparison of Flat Fields}\label{sec:comp}

Twilight sky flats are the most common form of flats applied to imaging
data at many observatories. They are typically assumed to represent the most constant
illumination of the detector and to give an image that removes
the individual pixel-to-pixel variations in response and any overall
illumination pattern imposed by the system optics. 
Of course, the observer has almost no control over the brightness
or color of these flats or any scattered light and color effects 
that are potentially problematic.

A standard test of a twilight sky flat is to compare it to an image
made from the median combination of a large number of deep nighttime
sky images (i.e. essentially a deep image of the sky without any stars). 
It is often presumed that if the twilight sky flat makes 
the nighttime sky image flat and free of any large scale 
residuals, then it must be a good approximation of the overall
illumination of the detector. The assumption is made that the illumination
of the detector by the nighttime sky is uniform and without
serious contaminants or significant illumination gradients.

\begin{figure}
\plotone{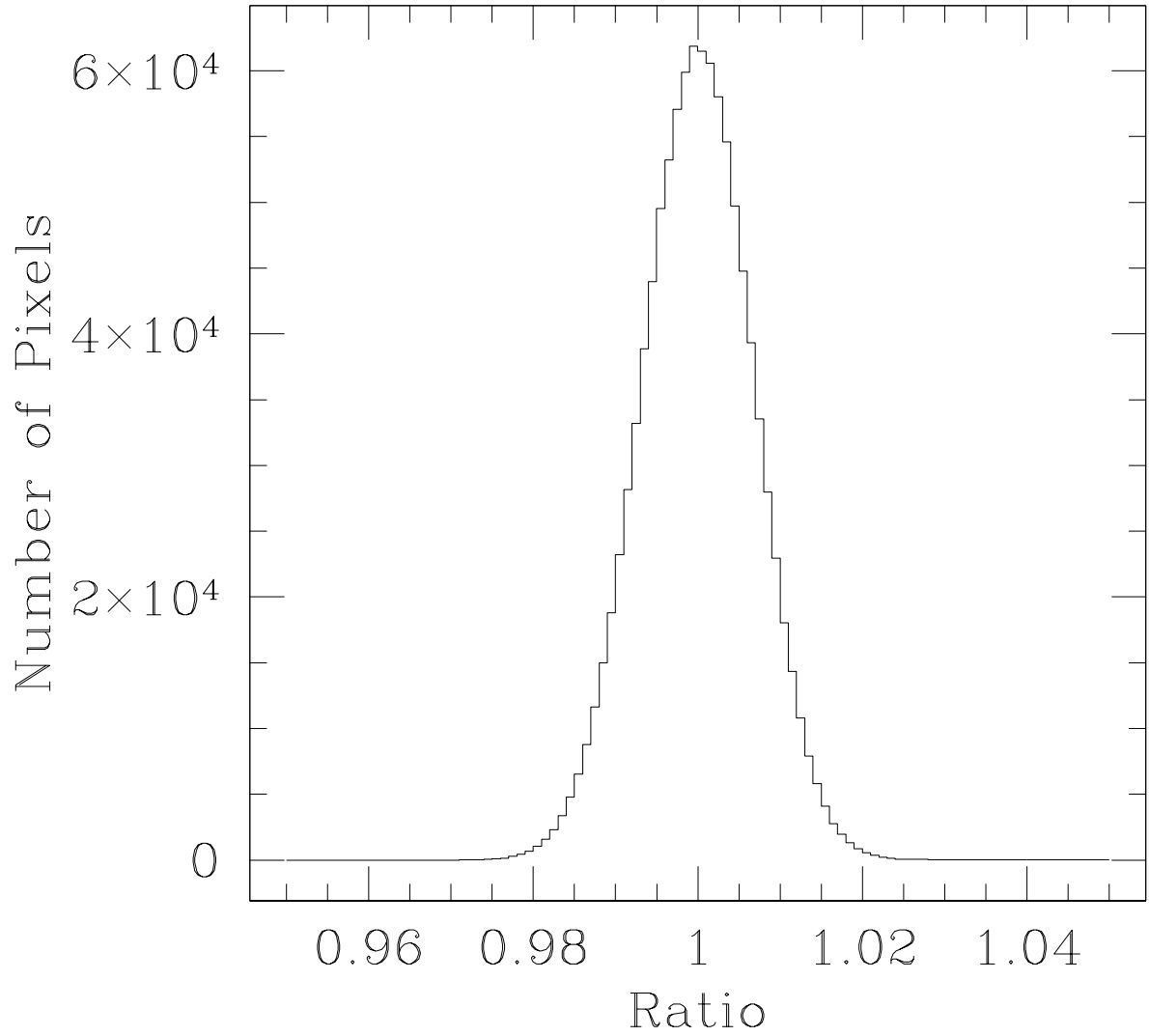}
\figcaption{
Histogram of values from the ratio of the twilight
sky flat to the nighttime sky flat. The histogram is normalized to 1, 
and has a standard deviation of 0.7\%
\label{fig:fig5}
}
\end{figure}

Indeed, when we ratio the twilight sky flat to the nighttime sky
flat described above we find a very uniform result. 
A histogram of the values in the
ratio is shown in Figure~\ref{fig:fig5}, which has been normalized to unity for
convenience. The standard deviation of the ratio is 0.7\%, roughly consistent
with the $S/N$ of the two individual flats (although dominated
by the $S/N$ of the nighttime sky flats). Further, there is 
no large scale structure in the ratio with an amplitude greater than
0.1\%.  This implies that the twilight sky flat and the nighttime sky flat 
are essentially the same; we will use only the twilight flat (since it has higher $S/N$) 
for the remainder of this analysis.  

\begin{figure}
\plotone{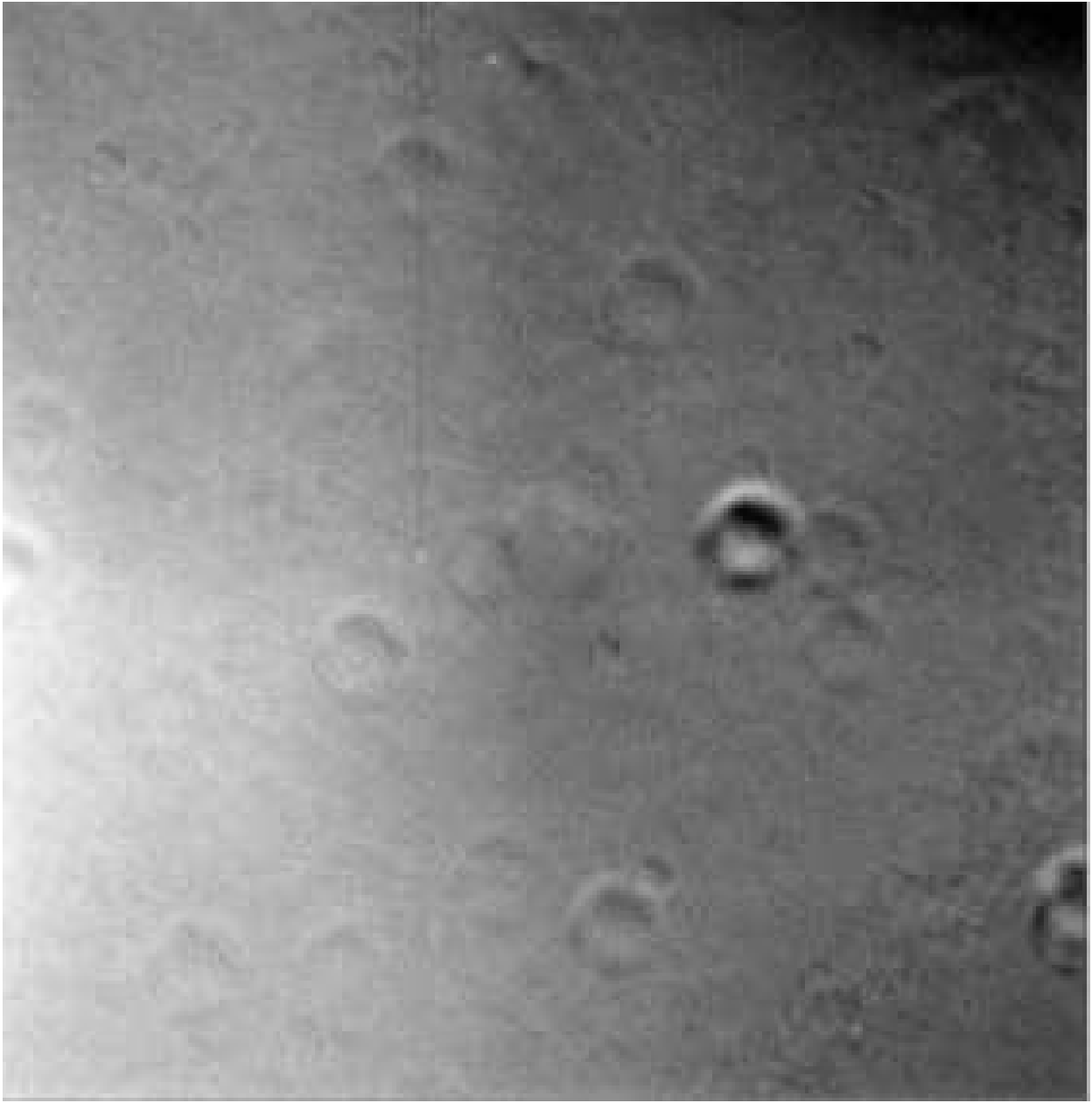}
\figcaption{
Ratio of a dome flat to the twilight sky flat. The grey scale used here 
ranges from roughly 0.97 
(black) to 1.03 (white).  The rings in the image are due to movement of dust particles 
on the optics of the system; the gradient across the image shows the 
different illumination pattern between the two flat fields.  
\label{fig:fig6}
}
\end{figure}

We tested the dome flats in a similar manner, creating a ratio of each of the three dome
flats to the twilight sky flat. In each case, there were obvious large
scale residuals in the result. These residuals were as large as 1.5\% in
amplitude over $\sim$100 pixel scales in the ratio; an example is 
shown in Figure~\ref{fig:fig6}. Some of the features seen in this image ratio
are clearly due to dust or other debris that
moved during the time between the dome flats and the twilight sky
flats; this material is most likely on the filter or cryostat window, 
both of which are near the focal plane. Even if these features
are ignored, there are other large scale structure and gradients
in the ratio of the dome flats to the twilight sky flat, suggesting
that there are significant differences in the illumination
of the detector between these two flat fields.
In Section~\ref{sec:stds} we discuss these differences in more detail.  

The three dome flat positions comprise three different combinations of 
telescope pointing and dome rotation.  For dome flat position \#1,
the telescope was pointed most directly at the flat field screen
and the screen was most nearly perpendicular to the optical axis of the 
telescope.  Dome flat position \#2 was nearly as well pointed, but the 
telescope was pointed slightly off the center of the screen; dome flat 
position \#3 was significantly off-center and the screen was not very perpendicular 
to the primary mirror.  

We compared the three dome flats by forming ratios of the flats obtained at each pointing.  
The standard deviation of the results was $\sim$2\%, larger than that expected from the estimated 
$S/N$ of the individual flats.  Indeed, there were obvious systematic 
features in the ratios.  These included features similar to those seen in 
Figure~\ref{fig:fig6}, although of somewhat smaller amplitude ($\sim$1\%).  This suggests 
again that dust or other obscuring material on the optics near the detector 
had moved slightly as the telescope was re-positioned.  There were also 
gradients seen in the ratios.  In the case of the ratio of dome flat \#1 to 
dome flat \#3, there was a gradient of $\sim$5\% diagonally across the 
detector.  A smaller gradient ($\sim$2\%) was seen in the ratio of dome flat 
\#1 to dome flat \#2; the direction of this gradient was nearly orthogonal to 
that seen in the ratio of dome flat \#1 to dome flat \#3.  

These tests would appear to suggest that the twilight sky flat is
the most accurate flat; the twilight sky flat certainly seems to make
the flattest and cleanest looking result when compared to the nighttime sky flat.
It is possible, however, that both the twilight sky flat and the nighttime sky flat 
simply have the same scattered light pattern.  If so, then the apparent 
cleanliness of their ratio does not indicate that the twilight sky represents 
the most accurate reproduction of the illumination of the detector by the 
science targets.  There is no direct way to determine whether small amounts 
of scattered light or other inappropriate light contaminates the 
twilight and nighttime sky flats equally.  

In general, image ratios do not contain sufficient information to properly evaluate 
the appropriateness of different flats.  Either the ratio appears ``clean'' 
(i.e. very flat, no small or large scale artifacts) or not.  If clean, then the 
ratio may simply indicate a common degree of contamination of the two frames.  
If not, then it may be impossible to tell which component of the ratio is problematic.  
We suspect that the twilight and nighttime sky flats are both contaminated by the same
amount and pattern of scattered light and that the illumination pattern is the same in both cases.  

\subsection{Tests of Flat Fields Using a Grid of Standard Stars}\label{sec:stds}

A well-designed flat field system should produce uniform
radiance of incident light at the entrance pupil of the telescope. 
Specifically, the flat field system should reproduce the pattern of incident 
light from science targets; both the spatial and angular patterns must be 
accurately simulated over the field of view of the detector.  
It is difficult to determine directly whether the system actually 
meets these requirements, since it would require
careful, calibrated determination of the radiance of light hitting 
all parts of the telescope primary mirror and from all incident angles of interest 
coming from the flat field system. 
Instead, we can use the
observation of many standard stars over the entire detector as 
a proxy for this difficult measurement.
The scatter in the derived photometric zero points of the standard stars 
can serve as a surrogate for a measurement of the uniformity of the 
entrance pupil illumination and defines a good  
measure of the appropriateness of the flat field. 

\begin{figure}
\plotone{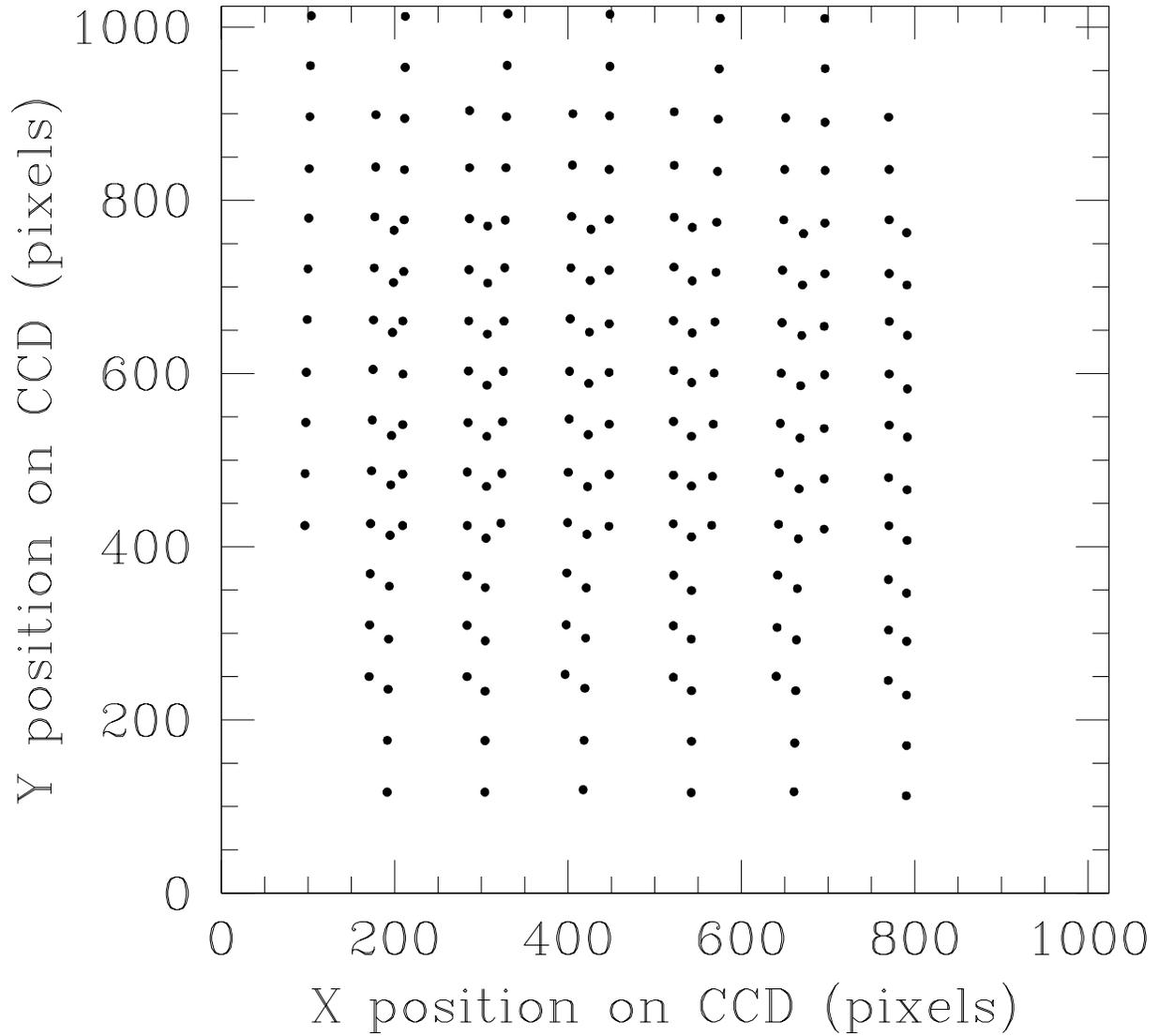}
\figcaption{
Positions of standard stars \citep[SA112-275, SA112-250, and SA112-223; see][]{l92} 
used to test the various flat fields.
\label{fig:fig7}
}
\end{figure}

We observed a field containing the 
standard stars SA112-275, SA112-250, and 
SA112-223 \citep[see][]{l92} a large number of times, 
moving the field center in a raster pattern between exposures. The 
pattern was a 12$\times$6 grid with a spacing of 
approximately 1 arcminute. Figure~\ref{fig:fig7} shows the positions of the standard 
stars over the entire set of images; roughly 75\% of the detector was covered. 
For some of the telescope pointings only
two of the standard stars were within the field-of-view of the CCD; on
a few others the star fell on a bad column or pixel. For these
reasons, there were a total of 205 useful measurements.
All observations were made using a 
standard broad-band Johnson V filter (the same as was used for the various flats) 
and had an exposure time of 5 seconds. 
The images were obtained at airmasses of 1.177 to 1.202. The night was clear 
and photometric, so the expected change in the extinction over the period 
of the observations is $<$0.002 mag.

The images of the standard stars were reduced in the usual manner using 
IRAF utilities, which consisted of overscan subtraction 
and application of a flat field.  
We compared four different flat fields applied to our data: a typical twilight sky 
flat and the three different dome flats described above. We also applied no
flat field to the images (i.e. only the bias was subtracted from the images
of the standard stars) to provide a baseline reference for the flats.

We derived photometric zero points from aperture photometry taken from 
all of the reduced images. 
A 12 arcsec aperture was used to measure each star; the sky
was estimated in an annulus of 16 to 24 arcsec in diameter around
each star. 
The seeing during the observations of the standard stars was $\sim$1.5 arcsec. 
Seeing variations were $<$5\% over the course of all the measurements, which 
should ensure that aperture related effects (i.e. a variable amount of the
stellar signal falling out of the aperture) are $<$0.001 mag.
Note that the colors of the three standard stars are very different 
($B-V=$0.45-1.21).  Since the combination of our filter and detector are not identical 
to the instrumentation used to determine the standard star magnitudes 
\citep[see][]{l92}, there is a color-dependent term that affects the 
absolute value of the derived zero point.  
This effect was clearly detected in our measurements; it was removed assuming a linear relation.  

To compare the relative quality of each of our five flat fields, 
we first simply derive the photometric zero point of each standard star as a baseline measurement.  
Table 2 shows the sample standard deviation of the zero point determination from each 
set of flattened images, including the set that had no flat field applied.  
Column 1 of Table 2 shows the standard deviation of all 205 determinations
of the zero point for each of the sets of images. In all cases applying
a flat field reduces the standard deviation; none of the flats seems to make the
determination of the zero point less accurate than doing nothing. The smallest
standard deviation is $\sim$0.66\%; this result was obtained using 
the dome flat in position \#1. For comparison, using a twilight sky flat
gave a standard deviation in the zero point determination of $\sim$0.93\%: better
than applying no flat at all, but worse than dome flat positions \#1 and \#2.  

A formal calculation of the expected signal from the stars suggests the
standard deviation should be $\sim$0.002 mag (if the flat fields had infinite $S/N$
and perfectly reproduced the telescope pupil).  
We suspect the difference between the theoretical standard deviation and what we 
measure is due to a combination of small effects: small instability in 
shutter timing, small photometric variations, and uncertainties in the 
brightness of stars. 

In an effort to find scattered light effects in any of the flat fields, we next 
fit various gradients to the photometric zero points and subtracted these 
small slopes from the data.  Columns 2-6 of Table 2 give these results 
as standard deviations of the zero points after this subtraction. 
We fit horizontal and vertical gradients, as well as gradients radiating from the center of 
the CCD and diagonal gradients from upper left to lower right and lower left to upper right.  
Each of these applied gradients makes an improvement in at least one of the flat fields.  
Note in particular that the twilight sky flat is improved significantly ($\sim$0.003 mag) 
by removal of a horizontal gradient.  

Dome flat position \#1 is least affected by removal of any gradient.  
For each of the gradients considered, the change in the standard deviation 
of the zero points is very small for dome flat \#1, typically $\lsim$0.001 mag.  
This suggests that dome flat \#1 has very 
good angular uniformity of illumination and is therefore the best flat field.  
Dome flat position \#2 is nearly as good as \#1.  Dome flat position \#3, 
which is pointed significantly off the center of the flat field screen, 
has significant gradient effects and is not as good a flat field as the 
previous two, demonstrating the importance of aligning the telescope and 
dome flat field screen in creating the flats.  

The very best flat field we obtained is the twilight sky flat 
after removal of the horizontal gradient.
In fact, the horizontal gradient seen in Figure~\ref{fig:fig6} can now be explained, and 
is wholly due to the twilight sky flat since the dome flat exhibits no gradient at all.  
It should be noted, however, that without the detailed analysis of hundreds of standard stars 
placed across the entire detector it would have been impossible to detect and remove this
gradient.  For all practical purposes, dome flat position \#1 
is the best flat field presented here.

We conclude that dome flat position \#1 is the best flat field for
determination of photometric zero points. This suggests it represents the
most even illumination of the telescope pupil and does an adequate job of removing
pixel-to-pixel variations (although the test presented here averages
out these small scale variations).


\section{Discussion \& Conclusions}\label{sec:conclude}

A well-constructed dome flat can have superior
performance to twilight sky flats, and indeed can flatten CCD data better than a twilight sky flat.  
Dome flats also have several practical advantages over sky flats,
most importantly that they are more easily and reliably obtained.
Twilight sky flats are prone to having lower than desirable $S/N$,
especially if the observer is inexperienced or the run is short; in particular, it is
difficult to obtain sky flats of many colors in one twilight period.  Twilight flats
are influenced by the presence of clouds in the field even though certain
nighttime observing programs may not be.  
Dome flats clearly have an advantage over sky flats in all these regards.  
Dome flats may be obtained at any time of the day, 
given a sufficiently dark dome environment.
They are very reproducible, and once one is assured that the dome 
flat is indeed superior to a
sky flat a dome flat field will be equally good night after night.  
Very high $S/N$ dome flats are straightforward to obtain by integrating 
as long as necessary and by taking a very large number of exposures.
These exposures may be obtained during the day to facilitate long integrations as
well as any and all bandpasses to be observed each night.

LEDs are an ideal illumination source for dome flat fields.  They are available 
in a wide range of colors and beam patterns; it is straightforward to combine 
a selection of LEDs
to illuminate a flat field screen of any size and in any bandpass.  In particular, 
LEDs of different colors can be combined to produce a flat field of 
almost any arbitrary color.  
LEDs are inexpensive and bright; not many units are required to 
illuminate a large screen.  They are easy to
use and to control with simple electronics, and do not have the time-variability issues
associated with incandescent illumination sources.  
Long lifetimes minimize time between replacement
and thereby reduce personnel commitments to maintain the system. 

The matching of color between flat fields and nighttime science observations 
may be a factor in the construction of flat field frames, 
and one which we do not address here.  
If this is the case, various LEDs may be carefully selected to produce any spectrum desired 
to illuminate a dome flat field screen.  

A similar system to that described here is currently in use
at the CTIO/SMARTS Consortium 1.3m telescope in Chile.
The system at this telescope also uses an array of LEDs
to illuminate a white screen.


\acknowledgements

The MDM Observatory and The Ohio State University 
Department of Astronomy provided support for the construction and installation of the flat field system.


%
%
\newpage

%
\clearpage


\begin{deluxetable}{cccccccc} 
\tablecolumns{8} 
\tablewidth{0pc} 
\tablecaption{LED Characteristics}
\tablehead{ 
\colhead{Color}    & \colhead{$\lambda_C$ \tablenotemark{a}} & \colhead{$\Delta\lambda$ \tablenotemark{b}} & \colhead{$V_F$ \tablenotemark{c}} & \colhead{$I_F$ \tablenotemark{d}} & \colhead{Viewing Angle} & \colhead{Manufacturer} & \colhead{Part Number} \\
\colhead{} & \colhead{(nm)} & \colhead{(nm)} & \colhead{(V)} & \colhead{(mA)} & \colhead{(degrees)} & \colhead{} & \colhead{} }
\startdata 
U & 370 & 23 & 3.9  & 10 & 110 & LEDtronics & BP200CUV1K-250 \\ 
B & 470 & 65 & 3.8 & 20 & 125 & LEDtronics & RL280TUB500-3.8 \\
V & 565 & 30 & 2.2 & 20 & 90  & Panasonic  & LN364GCP \\
R & 630 & 40 & 2.1 & 20 & 90  & Panasonic  & LN864RCP \\
I & 810 & 35 & 1.8 & 100 & 20 & Roithner   & ELD-810-525 \\

\enddata 

\tablenotetext{a}{Wavelength of peak brightness of spectral energy distribution}
\tablenotetext{b}{Full-width at half-maximum of the spectral energy distribution of the LED}
\tablenotetext{c}{Typical voltage drop across LED due to current flowing in the forward direction}
\tablenotetext{d}{Typical forward current needed for optimal light output of LED}

\end{deluxetable} 

\clearpage

\begin{deluxetable}{lcccccc}
\tablecolumns{7}
\tablewidth{0pc}
\tablecaption{Flat Field Comparisons: Standard Star Grid\label{tbl-1}}
\tablehead{
\colhead{}  & \multicolumn{6}{c}{Sample standard deviation of photometric zero point}\\
\colhead{}  & \multicolumn{6}{c}{after removal of gradient (mmag)\tablenotemark{a}}\\
& \colhead{no gradient} & \colhead{horizontal} & \colhead{vertical}
 & \colhead{radial from center} &\colhead{UL-LR} & \colhead{LL-UR} }
\startdata
                None & 15.1 & 9.6  & 13.9 & 14.6 & 9.8 & 14.7\\
        Twilight Sky & 9.3  & 6.0  & 9.4  & 9.1 & 8.3  & 8.0\\
Dome Flat Position 1 & 6.6  & 6.2  & 6.5  & 6.6  & 6.2  & 6.5\\
Dome Flat Position 2 & 8.3  & 7.8  & 7.1  & 8.0 & 6.9 & 8.0\\
Dome Flat Position 3 & 13.2 & 10.1 & 10.7 & 13.1 & 13.0 & 6.2\\
\enddata

\tablenotetext{a}{Standard deviation of 205 measurements; see text for a description of gradient forms.}

\end{deluxetable}

\end{document}